\documentclass[aps,prb,amsmath,showpacs,amssymb,twocolumn, groupedaddress]{revtex4-1}

\usepackage{graphicx}
\usepackage{xcolor}

\begin{document}


\title{Method for determining optimal supercell representation of interfaces}



\author{Daniele Stradi}
\email[]{daniele.stradi@quantumwise.com}
\affiliation{QuantumWise A/S, Fruebjergvej 3, PostBox 4, DK-2100 Copenahgen, Denmark}

\author{Line Jelver}
\affiliation{QuantumWise A/S, Fruebjergvej 3, PostBox 4, DK-2100 Copenahgen, Denmark}

\author{S\o ren Smidstrup}
\affiliation{QuantumWise A/S, Fruebjergvej 3, PostBox 4, DK-2100 Copenahgen, Denmark}

\author{Kurt Stokbro}
\affiliation{QuantumWise A/S, Fruebjergvej 3, PostBox 4, DK-2100 Copenahgen, Denmark}


\date{\today}

\begin{abstract}
The geometry and structure of an interface ultimately determines the behavior of devices at the nanoscale. We present a generic method to determine the possible lattice matches between two arbitrary surfaces and to calculate the strain of the corresponding matched interface. We apply this method to explore two relevant classes of interfaces for which accurate structural measurements of the interface are available: (i) the interface between pentacene crystals and the (111) surface of gold, and (ii) the interface between the semiconductor indium-arsenide and aluminum. For both systems, we demonstrate that the presented method predicts interface geometries in good agreement with those measured experimentally, which present nontrivial matching characteristics and would be difficult to guess without relying on automated structure-searching methods.  
\end{abstract}


\maketitle


\section{Introduction\label{Intro}}

As electronic devices shrink in size to reach nanoscale dimensions, interfaces between different materials become increasingly important in defining the device characteristics\citep{blom2013atomistic}. In many cases, it has been shown that the effect of the interface even dominates the device  properties\cite{dusastre2012interface}, leading to the concept that ``the interface is the device'' \citep{Kroemer2001}. In order to optimize the performance of a device it is therefore important to understand the properties of its interfaces. 

First principles modeling based on atomistic methods such as  Density Functional Theory (DFT) \citep{Martin2004} have become an important tool for simulating the properties of interfaces \citep{Stradi2016}. To be truly predictive, atomistic methods require an accurate model for the atomic-scale geometry of the interface. As these simulations typically use periodic boundary conditions in the direction parallel to the interface, a common supercell for the surfaces of the two crystals  forming the interface must be determined. However, typically the two crystals are not commensurate and finding a common supercell requires straining one of or both the surfaces. To accommodate the resulting strain, the two surfaces can also be rotated with respect to each other. However, for rotation angles preserving a high symmetry in the supercell, this has often the side-effect of increasing considerably its dimensions. Finding a supercell with low built-in strain and without an excessive number of atoms is therefore highly nontrivial. 

In this paper we present an algorithm which allows for an efficient and systematic search for common supercells between two crystalline surfaces. Given the optimized geometries of two surfaces forming the interface, the algorithm returns a list of all possible interface supercells by varying the interface strain and the rotation between the two surfaces. A related, but more simplistic method have been proposed in Ref.~\cite{lazic2015cellmatch}. Compared to  Ref.~\cite{lazic2015cellmatch} our method automatically tests all possible rotations of the two lattices and has been implemented into a graphical user interface, the Virtual NanoLab\cite{VNL2016.2}.

In the paper we show that this is not only a practical procedure for generating low strain supercells for atomic-scale simulations, but is a predictive tool for determining interface geometries in accordance with experimental data.  As a first example, we consider the interface between a pentacene crystal (PC) and the Au(111) surface, which has been widely studied both theoretically \citep{Ortega2011,Li-2009,Lee-2007,Lee-2005,Toyoda-2010,Saranya-2012} and experimentally \citep{Koch-2007,McDonald-2006,Schroeder-2002,Ihm-2006,France-2003,Kafer-2007,
Watkins-2002,Kang-2003,Soe-2009,Diao-2007,Liu-2010}. We show that the predicted geometries of a pentacene monolayer on Au(111) recover those observed experimentally. Using DFT, we calculate the ground state structure and energetics of these interfaces and find that they are thermodynamically more stable than those previously used in the literature. 

As a second example, we consider the interface between Al and InAs. This interface is relevant for studies on semiconductor nanowires (NWs) in which superconducting properties are introduced by proximity effect with a superconductor \citep{Deng2016,Albrecht2016,Higginbotham2015,Chang2015} and its structure has been recently resolved using high-resolution transmission electron microscopy (HR-TEM) \citep{Krogstrup2015}. 

The organisation of the paper is the following. In section~\ref{methods} we introduce the algorithm for matching the two crystal orientations with minimal strain. In section~\ref{ResultsAu} we first apply the method to determine the geometry of a pentacene overlayer on Au(111). In section~\ref{ResultsAl} we determine the structure of Al on InAs. Finally in section~\ref{conclusions} we conclude.
\section{Methods\label{methods}}

\begin{figure}
\centering
\includegraphics[width=0.45\textwidth]{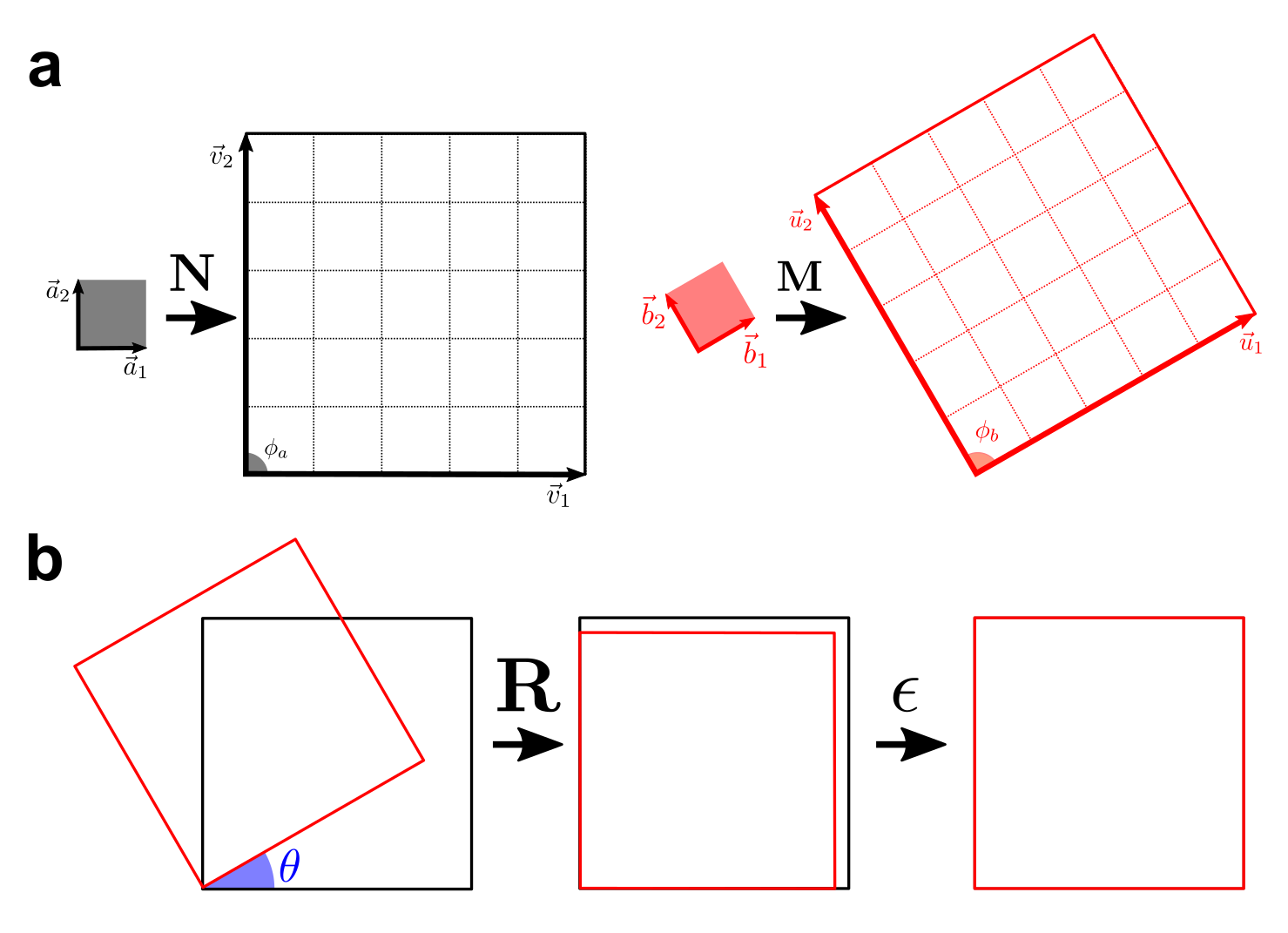}
\caption{\label{fig:method} Schematics of the method for determining the optimal supercell representation of an interface. (a) Two surface supercells with lattice vectors $(\vec{v}_1,\vec{v}_2)$ and $(\vec{u}_1,\vec{u}_2)$ are constructed from the surface unit cells with lattice vectors $(\vec{a}_1,\vec{a}_2)$ and $(\vec{b}_1,\vec{b}_2)$, respectively. (b) The two supercells are aligned and matched by first applying the rotation matrix $\mathbf{R}$ and then the strain tensor $\epsilon$.}
\end{figure}

\paragraph*{Algorithm for surface matching:}

A scheme of our algorithm for surface matching is shown in Fig. \ref{fig:method}. In order to obtain a low-strain interface structure, we systematically search through all possible 2D unit cells of the two surfaces forming the interface. Given two arbitrary surfaces $A$ and $B$ with primitive vectors $(\vec{a}_1, \vec{a}_2)$ and $(\vec{b}_1, \vec{b}_2)$, we generate the Bravais lattices of the possible surface supercells $A^*$ and $B^*$:

\begin{align}
(\vec{v}_1, \vec{v }_2) &= \mathbf{N} (\vec{a}_1, \vec{a}_2), \label{eq:supercells1}\\
(\vec{u}_1, \vec{u }_2) &= \mathbf{M} (\vec{b}_1, \vec{b}_2). \label{eq:supercells2}
\end{align}

In Eq. \ref{eq:supercells1}-\ref{eq:supercells2}, $\mathbf{N}$, $\mathbf{M}$ are $2\times2$ repetition matrices where the entries are integers
below two threshold values $N_\mathrm{max}$, $M_\mathrm{max}$, and ($\vec{v}_1, \vec{v }_2$), ($\vec{u}_1, \vec{u}_2$) are the Bravais lattice vectors of the resulting supercells. During the generation of the  supercells, we exclude equivalent lattices.

For each pair of supercells, we next determine a rotation matrix $\mathbf{R}$ which  rotates $B^*$ and aligns $\vec{u}_1$ with $\vec{v}_1$:

\begin{align}
\mathbf{R} = 
 \begin{bmatrix}
  \cos\theta &  -\sin\theta \\
  \sin\theta & \cos\theta
 \end{bmatrix}
\end{align}

\noindent where $\theta = |\phi_a - \phi_b|/2$, with $\phi_a = \angle (u_1,u_2)$ and $\phi_b = \angle (v_1,v_2)$, respectively. Finally, we  match the two supercells by defining a strain tensor $\epsilon$, which is applied to $B^*$ in order to match its Bravais lattice onto that of $A^*$. The resulting equation to match $A^*$ and $B^*$ reads: 

\begin{equation}
(\vec{v}_1, \vec{v }_2) = (1+\varepsilon) \mathbf{R} (\vec{u}_1, \vec{u }_2),
\label{eq:rotationstrain}
\end{equation}

\noindent with the individual components of the strain tensor $\varepsilon$ being: 

\begin{align}
\varepsilon_{xx} &= \left| \frac{v_{1,x}}{u_{1,x}} \right| - 1. \label{eq:straintensor1}\\
\varepsilon_{yy} &= \left| \frac{v_{2,y}}{u_{2,y}} \right|  - 1. \label{eq:straintensor2}\\
\varepsilon_{xy} &= \frac{1}{2} \frac{v_{2,x}-\frac{v_{1,x}}{u_{1,x}}u_{2,x}}{u_{2,y}}.\label{eq:straintensor3}
\end{align}

A similar procedure can also be be applied to strain $A^*$ and match it to $B^*$, or to strain equally both surfaces.

Straining one or both sides of the interface introduces and additional elastic contribution to the interface energetics. This contribution and its influence on the surface geometry varies considerably depending on the strength of the interaction between the overlayer and the substrate. For a substrate and a strained overlayer with cubic symmetry, we can write down the total energy per unit of area as\citep{carel1996computer}:

\begin{equation}
E = E^{int}+E^{surf}+
(\varepsilon_{xx}^2 C_{11} + 
\varepsilon_{xx} \varepsilon_{yy} C_{12}+
\frac{1}{2}\varepsilon_{xy}^2 C_{44}) t
\label{eq:energystrain1}
\end{equation}

where $E^{int}$ is the interface energy between the substrate and the overlayer, $E^{surf}$ is the energy of the free surface of the overlayer, $\varepsilon_{xx}, \varepsilon_{yy}, \varepsilon_{xy}$ have been defined in Eq. \ref{eq:straintensor1}-\ref{eq:straintensor3}, $C_{11}$, $C_{12}$, $C_{44}$ are the elastic constants of the overlayer material and $t$ is its thickness. The equation can be further simplified as

\begin{equation}
E = E^{int}+ E^{surf}+\frac{1}{2}\bar{\mathcal{E}}^2 C_{11} t
\label{eq:energystrain2}
\end{equation}

\noindent where

\begin{equation}
\bar{\mathcal{E}} = \sqrt{\varepsilon_{xx}^2+
\varepsilon_{yy}^2+ 
2 \frac{C_{12}}{C_{11}}  e_{xx}\varepsilon_{yy} +
\frac{C_{44}}{C_{11}}\varepsilon_{xy}^2},
\label{eq:energystrain3}
\end{equation}

Neglecting interactions between the interface and the overlayer free surface and strain effects, $E^{surf}$ will be independent of the interface geometry.  For metals and weakly interacting interfaces, we also expect the interface energy  $E^{int}$ to be  rather similar for different geometries, so that the contribution of the elastic energy will be dominant and will determine the stability trend of the different geometries. On the other hand, for interfaces between semiconductors there may be a varying number of bonds at the interface, depending on the precise overlay structure. Since binding energies for covalent bonds are typically in the range 1-2 eV, the overlayer may need to have a thickness above $\sim$2 nm before the elastic energy will dominate.

We have implemented the algorithm into the Virtual NanoLab\cite{VNL2016.2} and
the calculated matches are presented graphically as illustrated in Fig. \ref{fig:graph}, The plot shows the number of atoms in the supercell as function of the average strain $\bar{\varepsilon}$. For the average strain we for simplicity use,  $\bar{\varepsilon} =(|\varepsilon_{11}|+|\varepsilon_{22}|+ |\varepsilon_{12}|)/3$. The measure in Eq.~(\ref{eq:energystrain3}) gives slightly different orderings, but we found that orderings are basically similar and therefore selected the most simple option. The algorithm has been implemented to only test relevant vectors and with default values of $N_\mathrm{max},M_\mathrm{max} = 6$ a scan typically takes 1s and the simulation time scale as $N_\mathrm{max}^2$. In the next section we will apply the method to determine the structure of a Pentacene monolayer on Au(111) and the interface geometry of Al on top of InAs.

\section{Results\label{Results}}

\subsection{PC/Au(111)\label{ResultsAu}}

\begin{figure}
\centering
\includegraphics[width=0.45\textwidth]{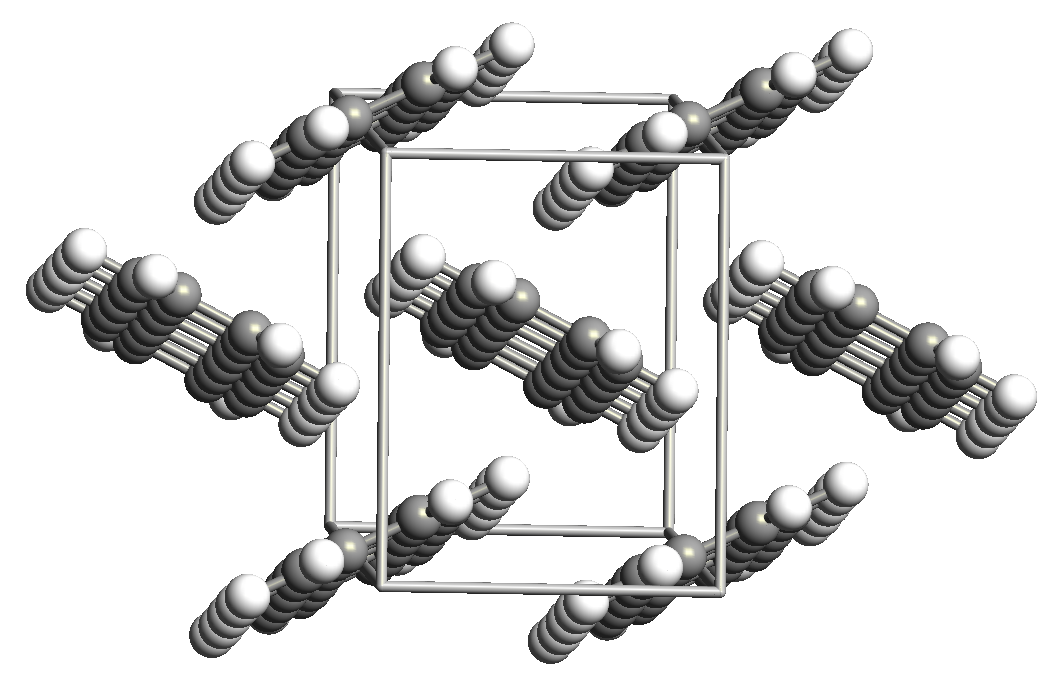}
\caption{\label{fig:pccrystal} Structure of the pentacene crystal.}
\end{figure}

\begin{figure}
\centering
\includegraphics[width=0.45\textwidth]{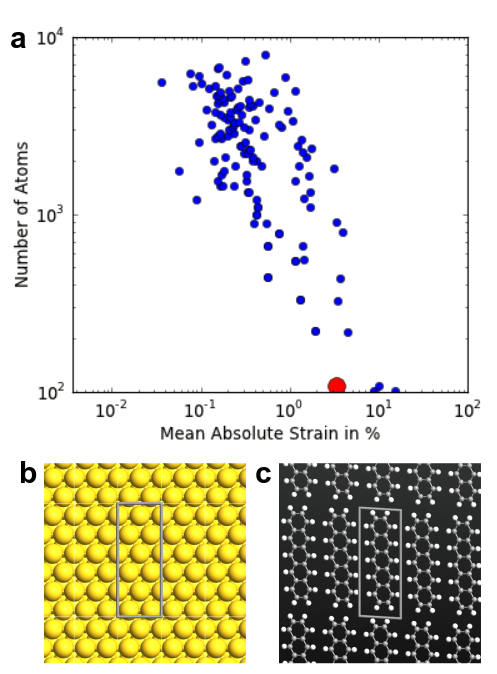}
\caption{\label{fig:graph} (a) Graph of the possible supercells generated for PC/Au(111) as a function of the mean absolute strain $\bar{\epsilon}$ in the pentacene crystal and the number of atoms in the supercell. (b,c) Au(111) and pentacene surface lattices associated with supercell II. The latter is highlighted by the red dot in (a).}
\end{figure}

\paragraph*{Computational details:} 

The DFT calculations for PC/Au(111) have been performed using the \textsc{Atomistix ToolKit} \citep{ATK2016.2}. The Kohn-Sham orbitals have been expanded in a linear combination of pseudo-atomic orbitals (PAOs) \citep{Junquera2001}. The electronic exchange-correlation (xc) energy has been described by using the generalized gradient approximation (GGA) and the Perdew-Burke 91 (PW91) xc-functional \citep{pw91}. We use this functional to compare with previous calculations\cite{Li-2009}, however, note that the GGA-PW91 xc-functional does not include van-der-Waals forces and it will therefore underestimate adsorption energies.
We have used a slab geometry with periodic boundary conditions parallel to the surface and mixed (Dirichlet+Neumann) boundary conditions in the direction normal to the surface, the latter allows for describing slabs with different workfunctions on the upper and lower surface. The Brillouin zone has been sampled using an $8 \times 3 \times 1$ Monkhorst-Pack \citep{Monkhorst1976} grid and a Fermi-Dirac occupation scheme with a broadening of $k_\mathrm{B}T = 25\ \mathrm{meV}$. Structural relaxations have been performed using a convergence threshold for the forces of $0.01\ \mathrm{eV}/\mathrm{\AA}$. During both the structural optimization and the evaluation of binding energies, the basis set superimposition error (BSSE) has been corrected using the counterpoise (CP) correction scheme\citep{Boys1970}. For the Au(111) surface, only the two uppermost layers were allowed to relax during the structural optimizations, while the atoms in the lowermost layers were kept frozen at their bulk position. 

For carbon we have used 21 orbitals per atom with s, p and d character and ranges up to 3.9 $\mathrm{\AA}$, while for hydrogen we have used 5 orbitals per atom with s and p character and ranges up to 4.2 $\mathrm{\AA}$. This basis set has been optimized to reproduce hydrogen and carbon dimer total energies \citep{Blum-2009}. Using this basis set, we obtain an adiabatic ionization energy for the individual pentacene  molecule (P1) $E_\mathrm{I} = 6.34\ \mathrm{eV}$ , in good agreement with the experimentally reported value $E_\mathrm{I}^\mathrm{exp} = 6.59\ \mathrm{eV}$  \citep{Gruhn-2002}. For gold, we have used an s, p, d basis set of ranges $2.7-3.6$ $\mathrm{\AA}$, with a total of 9 orbitals per atom. The calculated lattice constant for bulk Au using this basis is $a_\mathrm{Au} = 4.17\ \mathrm{\AA}$. Using a layer of gold ghost orbitals to get a better description of the isolated Au(111) surface \citep{GarciaGil2009}, we also obtain that the surface work function is $W_\mathrm{Au(111)} = 5.19\ \mathrm{eV}$. Both values are in good agreement with those obtained using similar computational parameters and a plane wave basis set \citep{Li-2009}.  

\paragraph*{Results:}

To construct the interface between the PC (see Fig.\ref{fig:pccrystal}) and Au(111), we have considered the PC unit cell according to the crystallographic parameters of Ref. \citenum{Sciefer-2006} (P-1: $\mathrm{a} = 5.985\ \mathrm{\AA}$, $\mathrm{b} = 7.596\ \mathrm{\AA}$, $\mathrm{c} = 15.6096\ \mathrm{\AA}$, $\mathrm{\alpha} = 81.25^\circ$, $\mathrm{\beta} = 86.56^\circ$, $\mathrm{\gamma} = 89.8^\circ$). The internal stress calculated in this unit cell is lower than $3\ \mathrm{meV}/\mathrm{\AA^3}$. 

We have then aligned the $\langle 010 \rangle$ direction of the PC unit cell along the normal to the Au(111) surface, and generated all possible interface supercells with $N_\mathrm{max},M_\mathrm{max} \leq 12$ by straining the PC lattice in the plane perpendicular to the $\langle 010 \rangle$ direction. In this case, we have used the experimental lattice parameter of the Au(111) surface. Fig. \ref{fig:graph}(a) shows a graph with the resulting possible supercells, sorted according to the mean absolute strain $\bar{\epsilon}$ and the number of atoms in the supercell. The lattices with lowest strain are listed in Table \ref{tab:strain}. It can be seen that supercell I, which has been used in earlier reports to model the PC/Au(111) surface \citep{Li-2009}, possesses a rather large internal strain. On the other hand, our method reveals the existence of other non-trivial  supercell arrangements which are associated with a much less strained PC lattice. In particular, among all the interfaces formed by a single PC(010) surface unit cell, the value of $\bar{\epsilon}$ is $67\%$ and $77\%$ lower for supercell II and III, respectively.  

\begin{table}
\caption{\label{tab:strain} Strain in the $\langle 010 \rangle$-oriented PC crystal to match
  Au(111). The first and second columns label each geometry by a roman number and list the supercell in the basis of the Au(111) bravais lattice. The third column list the number of PC(010) surface cells in the structure. $\varepsilon_{11}$,
  $\varepsilon_{22}$, $\varepsilon_{12}$ are the components of the
  strain tensor applied to the PC(010) surface cell in order to match the gold supercell. $\bar{\varepsilon} =(|\varepsilon_{11}|+
  |\varepsilon_{22}|+ |\varepsilon_{12}|)/3$, is the average strain.}
\begin{ruledtabular}
\begin{tabular}{ccccccc}
  Structure & Au(111) & $\#$PC & $\varepsilon_{11}$  & $\varepsilon_{22}$  &
  $\varepsilon_{12}$  & $\bar{\varepsilon}$ \\
      \cline{1-7}
I & (2,1;0,6)    &  1  &  -16.0 & 10.9 & 3.3  &    10.1 \\
II & (2,0;3,6)    & 1 &  -3.0  & -4.0 &  2.9 &    3.3  \\
III & (2,0;-3,6)   &  1  &  -6.4  & -0.54  &  0.0  &   2.3  \\
IV & (2,0;-2,13)   & 2  &   0.2  &  0.7  &  3.7  &  1.5  \\
V & (6,1;5,3)      & 4 & -0.8  &  -0.3 &  -0.5 & 0.5   \\
VI & (16,-1;9,-2)  & 12 & 0.0  &  0.1 &  0.1 & 0.1   \\
\end{tabular}
\end{ruledtabular}
\end{table}

\begin{figure}
\centering
\includegraphics[width=0.45\textwidth]{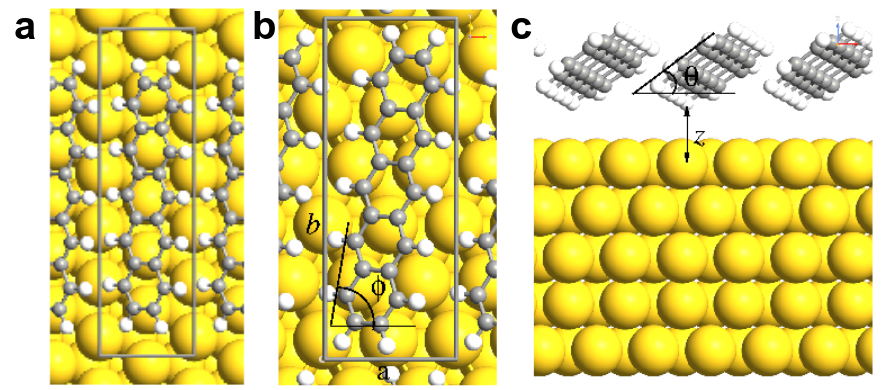}
\caption{\label{fig:pcau_structure} (a) Top view of supercell I. (b,c) Top and side views of supercell II. The structural parameters $a$, $b$, $\phi$, $\theta$ and $z$ are also shown (see the main text for a description of the parameters).}
\end{figure}

\begin{table*}
\caption{\label{pentacene-au111}
Calculated properties of the optimized geometries of PC/Au(111), for supercell I, II, III. The repetition of the Au(111) surface bravais vectors is given in parenthesis for each supercell. $E_b$ is the binding energy. $\Phi$ is the work function.
Reference calculation values  for supercell I obtained by Li
{\it et. al.}\cite{Li-2009, note-z} are given in the third column.
}
\begin{ruledtabular}
\begin{tabular}{cccccc}
 & I-(2,1;0,6) & Li {\it et. al.}\cite{Li-2009} & II-(2,0;3,6)  & III-(2,0;-3,6) & Exp.\\ \hline
      \cline{1-6}
$a$ (\AA) &  5.11   &5.11  &  5.90  & 5.90 & 5.64\cite{Schroeder-2002}
      5.76\cite{France-2003} 5.7\cite{Kafer-2007, Kang-2003} \\
$b$ (\AA)&  17.71  & 17.71  &  15.44  & 15.33 & 14.8\cite{Schroeder-2002}
      15.0\cite{France-2003} 15.5\cite{Kafer-2007, Kang-2003}\\
$z$ (\AA) &  3.35  & 3.1-3.5  &  3.18  & 3.17\\
$\theta\ (^\circ)$ & 40 & 38 & 36 & 34 & 43\cite{Ihm-2006} 31\cite{Kafer-2007}  \\
$\phi\ (^\circ)$ & 87 & 81 & 81 & 80 \\
$\Phi$  (eV) &   4.25  &  4.29  & 4.48 & 4.50 & 4.52 \cite{Schroeder-2002}
4.4\cite{Watkins-2002} 4.6\cite{Diao-2007}\\
$E_b$ (eV) & -0.29  & -0.16 & -0.42 & -0.42 & -1.14\cite{France-2003}  \\
\end{tabular}
\end{ruledtabular}
\end{table*}

To analyze the relationship between strain and adsorption properties in PC/Au(111), we have compared the optimized geometries of supercells I, II, and III (see Table~\ref{pentacene-au111}). For each optimized geometry we calculate the lattice vectors $a, b$ of the PC/Au(111) supercell, and the geometrical parameters of the PC crystal: the adsorption height $z$ and the polar and azimuthal adsorption angles $\theta$ and $\phi$, see Fig.~\ref{fig:pcau_structure}(b,c). Since supercell II and III have very similar properties, in the following we will only compare supercell I and II, see Fig. \ref{fig:pcau_structure} (a,b). The structural parameters obtained for supercell I are very similar to those obtained  using a plane wave basis set \citep{Li-2009}.  However, supercell II and III provides an overall better agreement with the available experimental data. In particular, the supercell lattice vectors $a, b$, and the azimuthal angle $\phi$ are closer to those measured experimentally.

In addition to the geometrical properties, we find that the calculated work function for supercell II is also in closer agreement with that measured experimentally, compared to that calculated for supercell I. Finally, the binding energy $E_\mathrm{b}$ calculated for supercell II is also larger compared to that calculated for supercell I. This indicates that the  supercell II and III, in addition to providing structural parameters in better agreement with those measured experimentally, lead to a structure which is thermodynamically more favorable.  We note that the discrepancy relative to the experimental value, is due to the neglect of van der Waals forces.

\subsection{Al/InAs\label{ResultsAl}}

\begin{table*}
\caption{\label{tab:strain_alinas} Predicted InAs/Al interfaces with the three lowermost strain for the two InAs surfaces considered. The second (fifth) and third (sixth) columns show the surface orientation of the InAs (Al) supercell and its structure in the basis of the primitive cell of the corresponding InAs (Al) surface. The fourth column shows the number of InAs surface cells in the structure.  
  $\varepsilon_{11}$,
  $\varepsilon_{22}$, $\varepsilon_{12}$ are the components of the
  strain tensor applied to the Al surface cell in order to match the gold supercell. $\bar{\varepsilon} =(|\varepsilon_{11}|+
  |\varepsilon_{22}|+ |\varepsilon_{12}|)/3$, is the average strain.}
\begin{ruledtabular}
\begin{tabular}{cccccccccc}
Structure & 
  InAs surface & 
  InAs & 
  $\#$InAs & 
  Al surface & 
  Al & 
  $\varepsilon_{11}$ &
  $\varepsilon_{22}$ &
  $\varepsilon_{12}$ &
  $\bar{\varepsilon}$ \\
      \cline{1-10}
I & $(1\bar{1}00)_\mathrm{WZ}$ & (2,0;0,1) & 2 & $(11\bar{2})$ &  (3,0;0,1) & -0.26 & -0.27 & 0 & 0.18 \\
II & $(1\bar{1}00)_\mathrm{WZ}$ &  (2,0;-1,2) & 4 & $(113)$ & (3,0;0,3) & -0.26 & -1.79 & 0 & 0.68 \\
III & $(1\bar{1}00)_\mathrm{WZ}$ & (0,-1;-3,0) & 3 & $(123)$ & (-1,1;2,1) & -0.27 & -2.06 & 0 & 0.78 \\

IV & $(111)\mathrm{B}_\mathrm{ZB}$ &  (2,2;0,2) & 4 & $(111)$ & (3,3;0,3) & -0.26 & -0.26 & 0 &  0.18 \\
V & $(111)\mathrm{B}_\mathrm{ZB}$ & (2,1;1,2) & 3 & $(111)$ & (3,2;1,3) & -2.06 & -2.06 & 0 &  1.37 \\
VI & $(111)\mathrm{B}_\mathrm{ZB}$ & (2,0;1,2) & 4 & $(11\bar{2})$ & (3,0;0,1) & -0.26 & 5.79 & 0 & 2.02 \\

\end{tabular}
\end{ruledtabular}
\end{table*}

\begin{figure}
\centering
\includegraphics[width=0.45\textwidth]{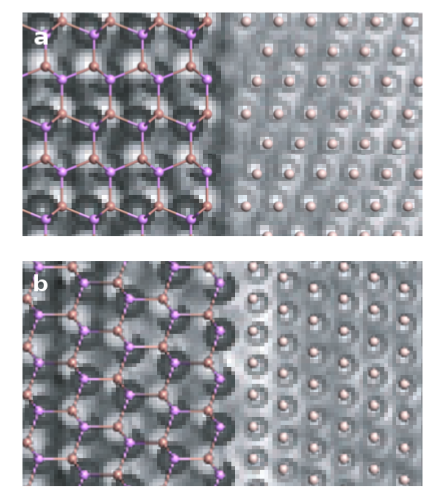}
\caption{\label{fig:tem}(a) Structure of the predicted InAs$(1\bar100)_\mathrm{WZ}$/Al$(11\bar{2})$ interface superimposed to the corresponding HR-TEM image measured experimentally. The In, As and Al atoms are shown as brown, violet and grey spheres. (b) Same as (a), but for the $(111)\mathrm{B}_\mathrm{ZB}$/Al$(111)$ interface.}
\end{figure}

Following recent experimental work on InAs NWs in which superconducting properties have been induced by the proximity effect with Al \citep{Deng2016,Albrecht2016,Higginbotham2015,Chang2015}, we have considered two surfaces of InAs: the $(1\bar{1}00)$ surface of the wurtzite phase (hereafter, $(1\bar{1}00)_\mathrm{WZ}$), and the $(111)\mathrm{B}$ surface of the zinc-blend phase (hereafter, $(111)\mathrm{B}_\mathrm{ZB}$). NWs with both these surfaces orientations have been grown experimentally, and it has been demonstrated that the precise orientation of the epitaxial Al overlayer depends on the exposed InAs surface \citep{Krogstrup2015}. 

For each of the two InAs surfaces, we have performed a scan over all Al$(mkl)$ surfaces with  $m,k,l \leq 3$. Subsequently, for each set of Miller indexes, we have generated all possible supercells with $N_\mathrm{max},M_\mathrm{max} \leq 6$ and with a maximum of four InAs surface cells,  by straining the Al lattice in the plane perpendicular to the interface.

For both the InAs surfaces considered, we have found that the Al surface which is predicted to have the lowest strain in the Al overlayer matches with that identified experimentally for thick ($t > 30\ \mathrm{nm}$) Al overlayers, see Table \ref{tab:strain_alinas}. In the case of InAs$(1\bar{1}00)_\mathrm{WZ}$, this corresponds to Al$(11\bar{2})$, whereas in the case of InAs$(111)\mathrm{B}_\mathrm{ZB}$, this corresponds to Al$(111)$. 

Another check on the accuracy of the method can be done by comparing the structures of the predicted InAs/Al interfaces with those measured experimentally. Fig. \ref{fig:tem} shows the predicted InAs/Al interfaces, superimposed to the measured HR-TEM images for each interface. It can be seen that, for both interfaces, the agreement between the structural model and the HR-TEM pattern is excellent. On the InAs side of the interfaces, regions of dark and bright contrast can be associated with In and As atoms, respectively, whereas on the Al side of the interfaces, the regions with bright contrast surrounded by a darker halo can be associated with Al atoms. 

\section{Conclusions\label{conclusions}}
In conclusion, we have presented a systematic and efficient method for determining a  supercell geometry  of the interface between two crystals. The method has been implemented into the Virtual NanoLab software. The method was applied to two metal-semiconductor systems, Au-Pentacene and InAs-Al interfaces. In both cases the method suggests interface geometries in good agreement with experimental data. For Au-Pentacene we illustrated that previous studies\cite{Li-2009}, which does not use a systematic approach for finding a supercell geometry of the interface have lower binding energies and are not in accordance with experimental data.

\begin{acknowledgments}
The  authors  acknowledges  support   from Innovation
Fund  Denmark under project 5189-00082B ``Atomic-scale modelling of interfaces'' and the Quantum Innovation Center (QUBIZ), Innovation Fund Denmark, and from the
European Commissions Seventh Framework Programme
(FP7/20072013),  Grant  Agreement  IIIV-MOS  Project
No.  619326. 
The authors would also like to thank Erik Johnson and Peter Krogstrup for providing the NW TEM images.  
\end{acknowledgments}

\bibliography{biblio}

\end{document}